\documentclass[aps,prd,twocolumn,preprintnumbers,superscriptaddress,floatfix]{revtex4}

\usepackage{multirow, graphicx,amssymb,url,mathrsfs,amsmath}
\usepackage{eucal,wrapfig,boxedminipage,setspace,subfigure}
\usepackage{amsxtra,amstext,latexsym,dsfont}

\def\IR{{\hbox{{\rm I}\kern-.2em\hbox{\rm R}}}}
\def\IB{{\hbox{{\rm I}\kern-.2em\hbox{\rm B}}}}
\def\IN{{\hbox{{\rm I}\kern-.2em\hbox{\rm N}}}}
\def\IC{\,\,{\hbox{{\rm I}\kern-.59em\hbox{\bf C}}}}
\def\IZ{{\hbox{{\rm Z}\kern-.4em\hbox{\rm Z}}}}
\def\IP{{\hbox{{\rm I}\kern-.2em\hbox{\rm P}}}}
\def\IH{{\hbox{{\rm I}\kern-.4em\hbox{\rm H}}}}
\def\ID{{\hbox{{\rm I}\kern-.2em\hbox{\rm D}}}}

\newcommand{\beq}{\begin{equation}}
\newcommand{\eeq}{\end{equation}}
\newcommand{\bea}{\begin{eqnarray}}
\newcommand{\eea}{\end{eqnarray}}

\begin{document}

\voffset 1cm

\newcommand\sect[1]{\emph{#1}---}

\title{Dynamic AdS/QCD and the Spectrum of Walking Gauge Theories}

\author{Timo Alho}
\email{timo.s.alho@jyu.fi}
\affiliation{Department of Physics, University of Jyv\"askyl\"a, P.O.Box 35, FIN-40014 Jyv\"askyl\"a, Finland}
\affiliation{Helsinki Institute of Physics, P.O.Box 64, FIN-00014 University of Helsinki, Finland}
\author{Nick Evans}
\email{evans@soton.ac.uk}
\affiliation{ STAG Research Centre \&  Physics and Astronomy, University of
Southampton, Southampton, SO17 1BJ, UK}
\author{Kimmo Tuominen}
\email{kimmo.i.tuominen@jyu.fi}
\affiliation{Department of Physics, University of Jyv\"askyl\"a, P.O.Box 35, FIN-40014 Jyv\"askyl\"a, Finland}
\affiliation{Helsinki Institute of Physics, P.O.Box 64, FIN-00014 University of Helsinki, Finland}

\begin{abstract}
We present a simple AdS/QCD model in which the formation of the chiral condensate is dynamically
determined. The gauge dynamics is input through the running of the quark bilinear's anomalous dimension, $\gamma$. The condensate provides a dynamically generated infra-red wall in the computation of mesonic bound state masses and decay constants. As an example, we use the model, with perturbative computations of the running of $\gamma$, to study SU(3) gauge theory with a continuous number of quark flavours, $N_f$.  We follow the behaviour of the spectrum as we approach the conformal window through a walking gauge theory regime. We show such walking theories display a BKT phase transition, with Miransky scaling, as one approaches the edge of the conformal window at the critical value of $N_f$. We show that these walking theories possess an enhanced quark condensate, a light Higgs like excitation and argue that the non-perturbative contribution to S falls to zero. We also study the deformation of the BKT transition when the quarks have a current mass which may be of use for understanding lattice simulations of walking theories.
\noindent

\end{abstract}

\maketitle

\newpage
\section{Introduction}

The AdS/QCD models \cite{Erlich:2005qh, DaRold:2005zs} 
provide a phenomenological holographic description of the QCD spectrum ($\pi$s, $\rho$s and $a$s)  and are able to capture some of the elements of the theory. The simplest model though does not describe the dynamics of the generation of the quark condensate, instead putting it in by hand. The decoupling of the
deep infra-red (IR) from the meson physics is also enacted by hand through a hard IR wall. In this paper we 
present an extension of these models, that includes all of this dynamics in one step. Our model is not a 
holographic description of the full theory since one must include the gauge dynamics via an assumed form for the running of the anomalous dimension of the quark anti-quark bilinear, $\gamma$. Given this input, %though the chiral condensate, 
the IR dynamical quark mass and the usual AdS/QCD spectrum are predictions. In addition the model includes the scalar meson (which in QCD is associated with the physical $f_0$). 

The model we present can be viewed as a natural evolution of work on D3/probe-D7 models \cite{Karch:2002sh} of chiral symmetry breaking \cite{Babington:2003vm} using the AdS/CFT correspondence \cite{Maldacena:1997re}, which we briefly review in the appendix. There has been work on trying to describe the full gauge dynamics of chiral symmetry breaking theories in this system \cite{Babington:2003vm, Filev:2007gb}, but it remains very hard to find true supergravity solutions of sufficiently complex dynamics. A phenomenological approach in this setup is to impose running on various AdS fields (such as the dilaton, that represents the gauge coupling) without worrying about back-reaction on the geometry \cite{Evans:2011eu}. Recently though we have shown that these models reduce in the quenched quark sector to dialling the renormalization group flow of the quark anti-quark bilinear operator \cite{Alvares:2012kr}. In the model we present in this paper we extract that element of the more complex models. One advantage of the link to the more rigorous string constructions is that the implementation of the IR wall in the model can be cleanly linked to the formation of the quark condensate.  

In \cite{Evans:2013vca} we investigated the core dynamics of this model for the quark condensate and the 
scalar meson. Here we 
couple that dynamical ``engine" to the AdS/QCD model to provide a wider description of the spectrum. The model predictions are dependent on the assumed form of $\gamma$,
%the model predictions are dependent on 
% if you believe you know that running. 
and therefore the power of the model is not so much for two or three flavour QCD but in allowing us to compare the behaviour of the spectrum of a set of theories as the form of the running changes. Such an application is to the $N_f$ dependence of the spectrum in SU($N_c$) QCD. Other instructive holographic descriptions of this physics can be found in \cite{Hong:2006si}-\cite{Goykhman:2012az}

For a theory with quarks in the fundamental representation asymptotic freedom sets in when 
$N_f<11/2 N_c$. Immediately below that point, at least at large $N_c$, the two loop beta function enforces a perturbative infra-red (IR) fixed point 
\cite{Caswell:1974gg,Banks:1981nn}. The fixed point behaviour is expected to persist into the non perturbative regime as $N_f$ is further reduced \cite{Appelquist:1996dq}. At some critical value of the number of flavours, $N_f^c$, the coupling is expected to be strong enough to trigger chiral symmetry breaking by the formation of a quark anti-quark condensate. 
The critical value, $N_f^c$, i.e. the lower boundary of the conformal window, can be estimated in a variety of ways \cite{Appelquist:1996dq}-\cite{Iwasaki:2003de}. 
Different semi-analytic methods typically yield the chiral transition to occur below $N_f^c\simeq  4N_c$ for fundamental fermion flavours. 
The chiral phase transition at the lower edge of the conformal window may give way to a regime of walking dynamics directly below $N_f^c$ \cite{Holdom:1981rm}.  For walking theories, there is expected to be a long energy range in which the coupling barely runs before tripping through the critical coupling value in the deep IR. Such theories display a tuned gap between the value of the quark condensate and the pion decay constant $f_\pi$, since the quark bilinear has a significant anomalous dimension, $\gamma$, over a large running regime. They are 
of interest phenomenologically for technicolor models of electroweak symmetry breaking 
\cite{Weinberg:1975gm, Susskind:1978ms} because flavour changing neutral currents are suppressed in extended technicolor models, and flavour physics is decoupled from the electroweak scale \cite{Holdom:1981rm}.
The walking dynamics may also suppress the contributions of the techni-quarks to the electroweak oblique corrections, in particular to the S parameter \cite{Sundrum:1991rf,Appelquist:1998xf}.  Walking theories may also possess a parametrically light bound state, a pseudo-Goldstone boson of the breaking of dilatation symmetry by the quark condensate \cite{Yamawaki:1985zg,Bando:1986bg,Hong:2004td,Dietrich:2005jn}.  
It is believed that in walking theories the intrinsic scale falls towards zero exponentially with $N_f^c-N_f$ as the conformal window is approached from below (known as Miransky scaling \cite{Miransky:1996pd} or a holographic BKT transition \cite{Kaplan:2009kr}).

In this paper we will take our running from the two loop perturbative result for the SU(3) gauge theory with a continuous flavour parameter $N_f$ allowing us to see all of the above structure and match the spectrum to that observed in QCD. 
%The generic behaviours with the different strengths of fixed points and running are expected to hold over a %wider range of $N_c, N_f$ though.
Our simple AdS model will display all characteristics of walking theories as we approach the conformal window. Holographically the quark condensate becomes non-zero when the scalar describing it in AdS suffers an instability - this occurs when its mass passes through the Breitenlohner-Freedman bound of $m^2=-4$ \cite{Breitenlohner:1982jf}. Using the usual AdS/CFT dictionary ($m^2= \Delta(\Delta-4)$) this corresponds to the point $\gamma=1$. Our assumption is that, as one decreases the continuous variable $N_f$ within the conformal window, the IR value of the scalar mass smoothly interpolates through $-4$. We show this leads to a continuous transition with BKT scaling. The spectrum, which as a whole falls to zero mass as one approaches the critical value, $N_f^c$, displays a relatively light sigma meson close to the critical value (which one might hope to link to the observed light Higgs mass \cite{atlas:2012gk,cms:2012gu}).  The mass splitting between the vector and axial vector mesons in the model is determined by the gauge coupling constant $\kappa$ in the 
five dimensional model, which is a free parameter. To fix it we make use of the observed splitting in QCD. For this reason we will study the $N_c=3$ theory and fix $\kappa$ at $N_f=2$. We will treat $N_f$ as a continuous parameter even at $N_c=3$ since the behaviour of the fixed points in the two loop QCD $\beta$ function show the same broad features with $N_f$ at all $N_c$. We believe all the features we observe will be present at any value of $N_c$ as one approaches the transition to the conformal window - of course formally only at the Veneziano limit (i.e. taking both $N_c$ and $N_f$ infinite with $x_f\equiv N_c/N_f$ fixed)  can one treat $N_f$ as a truly continuous parameter.  As one moves to higher $N_f$, the dependence of  $\kappa$ on $N_f$ is crucial - we assume that as we approach the {\it continuous} chiral symmetry restoration transition at $N_f^c$ the mass difference must fall to zero so that axial-vector symmetry is smoothly restored. Given this assumption it naturally follows that the contribution of vector and axial vector mesons to the S parameter falls to zero at the symmetry restoration point.

Finally we study the effect of including an explicit quark mass so the quarks are blind to the deep IR of the theory. The result is a deformation of the BKT type scaling. When the scale where the theory violates the BF bound is larger than the hard quark mass the theory behaves as the massless theory. The dynamics then smoothly moves over, as the quark mass becomes larger than the dynamical scale, to a regime where all mass scales in the theory (bound state masses, decay constants and so forth) scale as simple powers of the hard mass, as determined by dimensional analysis. These results will potentially be of use to guide lattice simulations of walking dynamics where the chiral limit is hard to achieve. 
\bigskip 

\section{Dynamic AdS/QCD}

The essential dynamics of our model is encoded into a field $X$ of mass dimension one. The modulus
of this field describes
%The modulus of the dimension 1 field $X$ describes 
the quark condensate degree of freedom. Fluctuations in $|X|$ around its vacuum configurations will describe the scalar meson. The $\pi$ fields are the phase of $X$
\begin{equation} X = L(\rho)  ~ e^{2 i \pi^a T^a} .
\end{equation}
Here $\rho$ is the holographic coordinate ($\rho=0$ is the IR, $\rho \rightarrow \infty$ the ultraviolet (UV)), 
and $|X|=L$ enters into the effective radial coordinate in the space, i.e. $r^2 = \rho^2 + |X|^2$. This is how the quark condensate will generate a soft IR wall: when $L$ is nonzero the theory will exclude the deep IR at $r=0$. This implementation is taken directly from the D3/probe-D7 model \cite{Karch:2002sh} where $L$ is the embedding of the D7 brane in the AdS spacetime. Fluctuations on the brane then see the pulled back metric on the D7 world volume. We briefly review the 
D3/D7 case in the Appendix.

We work with the five dimensional metric %for contractions of the space-time indices is given by
\begin{equation} 
ds^2 =  { d \rho^2 \over (\rho^2 + |X|^2)} +  (\rho^2 + |X|^2) dx^2, 
\end{equation}
which will be used for contractions of the space-time indices.
The five dimensional action of our effective holographic theory is
\bea
S & = & \int d^4x~ d \rho\, {\rm{Tr}}\, \rho^3 
\left[  {1 \over \rho^2 + |X|^2} |D X|^2 \right. \nonumber \\ 
&& \left.+  {\Delta m^2 \over \rho^2} |X|^2   + {1 \over 2 \kappa^2} (F_V^2 + F_A^2) \right], 
\label{daq}
\eea
where $F_V$ and $F_A$ are vector fields that will describe the vector ($V$) and axial ($A$) mesons.
Note that we have not written the $\sqrt{-g}$ factor in the metric as $r^3$ but just $\rho^3$. Again, this is driven by the D7 probe action in which this factor is $\rho^3$;  maintaining this form is crucial to correctly implementing the soft wall behaviour. 
%The choice of $\Delta m^2(r)$ will then be used to introduce an energy scale dependent mass for $L$ i.e. an anomalous dimension for $\bar{q}q$. 
Finally $\kappa$ is a constant that will determine the $V -A$ mass splitting; we will fix its value and $N_f$ dependence in our model below.

%\subsection{UV External V, A $\&$ S Currents}

The normalizations are determined by matching to the gauge theory in the UV of the theory. External currents are associated with the non-normalizable modes of the fields in AdS. In the UV we expect 
$|X| \sim 0$ and we can solve
the equations of motion for the scalar, $L= K_S(\rho) e^{-i q.x}$, vector $V^\mu= \epsilon^\mu K_V(\rho) e^{-i q.x}$, and  axial $A^\mu= \epsilon^\mu K_A(\rho) e^{-i q.x}$ fields. Each satisfies the same equation
\begin{equation}  \label{thing}
\partial_\rho [ \rho^2 \partial_\rho K] - {q^2 \over \rho} K= 0\,. \end{equation}
The UV solution  is
\begin{equation} \label{Ks}
K_i = N_i \left( 1 + {q^2 \over 4 \rho^2} \ln (q^2/ \rho^2) \right),\quad (i=S,V,A),
\end{equation}
where $N_i$ are normalization constants that are not fixed by the linearized equation of motion.
Substituting these solutions back into the action gives the scalar correlator $\Pi_{SS}$, the vector correlator $\Pi_{VV}$ and axial vector correlator $\Pi_{AA}$. Performing the usual matching to the UV gauge theory requires us to set
\begin{equation} N_S^2 = {N_c N_f \over 24 \pi^2 }, \hspace{0.5cm} N_V^2 = N_A^2 = {\kappa^2 N_c N_f \over 24 \pi^2 }.
\end{equation}
These choices should be compared to those in \cite{Erlich:2005qh, DaRold:2005zs}: In \cite{DaRold:2005zs}  the entire action is multiplied by the factor of $N_S^2$ so that the normalizations of the fields can be set to one. Then in \cite{DaRold:2005zs} the choice $\kappa=1$ is imposed so the scalar and vector correlators look symmetric. In \cite{Erlich:2005qh}  $\kappa$ is taken to equal our choice of $1/N_S^2$ making the vector and axial normalizations unity (but the scalar normalization is not one). The key point here is that $\kappa$ is not determined by the matching we have performed. 

In the next section we will derive the equations for the vacuum profile of the scalar $X$, the
equations of motion for the mesons and the formulas for evaluating the decay constants
%one should then re-compute the full solutions to the equations of motion with the boundary condition $K_i'(0)=0$ and 
subject to the normalization conditions on external currents we have just determined in the UV. 

\subsection{Vacuum Structure and Fluctuations}

Let us first consider the vacuum structure of the theory by setting all fields except $|X|=L$ to zero. We further assume that $L$ will have no dependence on the $x$ coordinates. The action for $L$  is given by
\begin{equation} \label{act} S  =  \int d^4x~ d \rho ~  \rho^3 \left[   (\partial_\rho  L)^2 +  \Delta m^2 {L^2  \over \rho^2 }   \right].
\end{equation}
Now if we re-write $L =  \rho \phi $ and integrate the first term by parts we arrive at
\begin{equation}  S  =  \int d^4x~ d \rho ~ ( \rho^5  (\partial_\rho  \phi)^2 +  \rho^3 (-3+ \Delta m^2) \phi^2  )\,,
\end{equation}
which is the form for a canonical scalar in AdS$_5$. The usual AdS relation between the scalar mass squared and the dimension of the field theory operator applies ($m^2 = \Delta (\Delta-4)$).   If $\Delta m^2 =0 $ then the scalar describes a dimension 3 operator and dimension 1 source as is required for it to represent $\bar{q} q$ and the quark mass $m$. That is, in the UV the solution for the $\phi$ equation of motion is $\phi = m/\rho + \bar{q}q/\rho^3$.

The Euler-Lagrange equation for the determination of $L$, in the case of a constant $\Delta m^2$, is 
\begin{equation} \label{embedeqn}
\partial_\rho[ \rho^3 \partial_\rho L]  - \rho \Delta m^2 L  = 0\,. \end{equation}
We have introduced $\Delta m^2$ in the full Lagrangian of the model in the minimal way consistent with changing the mass squared in the linearized regime. We can now ansatz an $r$ dependent $\Delta m^2$ to describe the running of the dimension of $\bar{q}q$. If the mass squared of the scalar violates the BF bound of -4 ($\Delta m^2=-1$) then we expect the scalar field $L$ to become unstable and settle to some non-zero value. 
If $\Delta m^2$ depends on $L$ then there is an additional term $- \rho L^2 m^{2\prime}(L)$ in the above equation of motion. At the level of the equation of motion this is an effective contribution to the running of the anomalous dimension $\gamma$ that depends on the gradient of the rate of running in the gauge theory. At one loop in the gauge theory there is no such term and so we will neglect this term, effectively imposing the RG running of $\Delta m^2$ only at the level of the equations of motion. Of course it is entirely appropriate to drop the term when the rate of running is small, which will be the case near the edge of the conformal window in the model below. For this reason including this term or otherwise makes no effect on the behaviour of the model as one approaches the phase transition from the chiral symmetry breaking phase to the conformal window, which is our main interest here.  

The solution for $L(\rho)$ can  be found numerically by shooting from $\rho=0$ with the IR boundary condition $L'(0)=0$. Adjusting the value of $L(0)$, the effective IR quark mass, $m_q$, one can find the regular flow that has $L(\infty) =m_q$ to describe a particular current quark mass. Until our final section we will set $m_q=0$ and study the chiral limit. The first two terms generate the dynamics %we have described 
with $L$ becoming unstable if $\Delta m^2 <-1$ over some range of $\rho$. 

%\subsection{$\sigma/f_0$ meson}

We next compute the scalar $\bar{q}q$ meson masses of our model. We look
for space-time dependent excitations on top of the vacuum configuration, $L_0$ ie $|X| = L_0 + \delta(\rho) e^{-i q.x}$,  $q^2=-M^2$. The equation of motion for $\delta$ is, linearizing (\ref{embedeqn}),
\begin{equation} \label{deleom} \begin{array}{c}\partial_\rho( \rho^3 \delta' ) - \Delta m^2 \rho \delta -   \rho L_0 \delta \left. \frac{\partial \Delta m^2}{\partial L} \right|_{L_0} \\ \\ 
+ M^2 R^4 \frac{\rho^3}{(L_0^2 + \rho^2)^2} \delta  = 0\,. \end{array} \end{equation}
We seek solutions with, in the UV, asymptotics of $\delta=1/\rho^2$ and with $\delta'(0)=0$ in the IR, giving a discrete meson spectrum. 

We must normalize $\delta$ so that the kinetic term of the $\sigma$ meson is canonical i.e.
\begin{equation} \int d \rho {\rho^3 \over (\rho^2 + L^2)^2} \delta^2 = 1\,. \end{equation}

The scalar meson decay constant can be found using the solutions for the normalizable and non-normalizable wave functions. We concentrate on the action term (after integration by parts)
\begin{equation}
S   = \int d^4x~ d \rho ~~ \partial_\rho (- \rho^3 \partial_\rho L) L\,.
\end{equation}
We substitute in the normalized solution $\delta$ and the external non-normalizable scalar function $K_S$
at $q^2=0$ with normalization $N_S$ to obtain the dimension one decay constant $f_S$ as 
\begin{equation}
f_S^2 =  \int d \rho \partial_\rho (- \rho^3 \partial_\rho \delta)  K_S(q^2=0)\,.
\end{equation}

%\subsection{$\rho$ meson}

The vector meson spectrum is determined from the normalizable solution of the equation of motion for the spatial pieces of the vector gauge field $V_{\mu \perp} = \epsilon^\mu V(\rho) e^{-i q.x}$ with $q^2=-M^2$. The appropriate equation is
\begin{equation} \label{vv}  \partial_\rho \left[ \rho^3 \partial_\rho V \right] + {\rho^3 M^2 \over (L_0^2 + \rho^2)^2} V = 0\,. \end{equation}
We again impose $V'(0)=0$ in the IR and require in the UV that $V\sim c/\rho^2$. To fix $c$ we normalize the wave functions such that the vector meson kinetic term is canonical
\begin{equation}  \int d \rho {\rho^3  \over \kappa^2 (\rho^2 + L_0^2)^2} V^2 = 1\,. \end{equation}

The vector meson decay constant is given by substituting the solution back into the action and determining the coupling to an external $q^2=0$ vector current with wave function $K_V$. We have for the dimension one $f_V$
\begin{equation} f_V^2 = \int d \rho {1 \over \kappa^2} \partial_\rho \left[- \rho^3 \partial_\rho V\right] K_V(q^2=0)\,.
\label{rhodecay}
\end{equation} 
Note here that the factors of $\kappa$ cancel against those in the normalizations of $V, K_V$ and the result is $\kappa$ independent.

%\subsection{$a$ meson}

The axial meson spectrum is determined from the equation of motion for the spatial pieces of the axial-vector gauge field. In the $A_z=0$ gauge we write $A_\mu= A_{\mu \perp} + \partial_\mu \phi$. 
The appropriate equation with $A_{\mu \perp} = \epsilon^\mu A(\rho) e^{-i q.x}$ with $q^2=-M^2$ is
\begin{equation}  \label{aa}  \partial_\rho \left[ \rho^3 \partial_\rho A \right]   - \kappa^2 {L_0^2 \rho^3 \over  (L_0^2 + \rho^2)^2} A + {\rho^3 M^2 \over (L_0^2 + \rho^2)^2} A = 0\,. \end{equation}
The asymptotic behaviour, boundary condition at $\rho=0$ and the normalization of $A(\rho)$ are same as those for $V(\rho)$.
%We again impose $A'(0)=0$ and require in the UV that $A=1/\rho^2$. The field $A$ is normalized similarly to $V$. 
The axial meson decay constant is given by (\ref{rhodecay}) with replacement $V\rightarrow A$.
%\begin{equation} F_a = \int d \rho {1 \over \kappa^2} \partial_\rho \left[- \rho^3 \partial_\rho A\right] K_A(q^2=0)
%\end{equation} 

%\subsection{$\pi$ meson}

 The $\phi^a$ and $\pi^a$ equations are mixed and of the form
\begin{equation}
\partial_\rho [ \rho^3 \partial_\rho \phi^a] - \kappa^2 {\rho^3 L^2 \over (\rho^2 + L^2)^2} ( \pi^a - \phi^a) = 0\,,
\end{equation}
\begin{equation}
- q^2  \partial_\rho \phi^a + \kappa^2 L^2 \partial_\rho \pi^a = 0\,. \end{equation}
Note that there is always a solution of these equations where $q^2=0$ and $\phi^a= \pi^a=$ a constant. The linearized field in the action associated with the pion is $X \sim 2 \pi i L_0$. Only in the case where the quark mass is zero and $L \sim c/ \rho^2$ asymptotically is this a normalizable fluctuation corresponding to a physical state in the field theory - it is the massless pion. When the quark mass is non-zero this state is a flat direction of the theory only if one makes a spurious transformation on $\bar{q}_L q_R$ and $m$ simultaneously - i.e. it is not a physical state in the spectrum.  

The pion decay constant can be extracted from the expectation that $\Pi_{AA} = f_\pi^2$. From the $f_A$ kinetic term with two external (non-normalizable) axial currents at $Q^2=0$ we obtain
\begin{equation} f_\pi^2 = \int d \rho {1 \over \kappa^2}  \partial_\rho \left[  \rho^3 \partial_\rho K_A(q^2=0)\right] K_A(q^2=0)\,.
\end{equation} 
\newpage

\section{Two Loop Running Inspired Model}

%\subsection{The model definitions}
To enact a realization of our model we must choose how the anomalous dimension of 
$\langle\bar{q}q\rangle$ runs with the energy scale through the function $\Delta m^2$. We will choose to look at the $N_f$ dependence of SU($N_c$) gauge dynamics to study the implications for the spectrum in the walking gauge theories expected to lie on the edge of the transition from the chiral symmetry breaking phase to the conformal window. We will use the perturbative running from SU($N_c$) gauge theories with $N_f$ flavours since the two loop results display a conformal window.  

The two loop running of the gauge coupling in QCD is given by
\begin{equation} 
\mu { d \alpha \over d \mu} = - b_0 \alpha^2 - b_1 \alpha^3,
\end{equation}
where
\begin{equation} b_0 = {1 \over 6 \pi} (11 N_c - 2N_F), \end{equation}
and
\begin{equation} b_1 = {1 \over 24 \pi^2} \left(34 N_c^2 - 10 N_c N_f - 3 {N_c^2 -1 \over N_c} N_F \right) .\end{equation}
Asymptotic freedom is present provided $N_f < 11/2 N_c$. There is an IR fixed point with value
\begin{equation} \alpha_* = -b_0/b_1\,, \end{equation}
which rises to infinity at $N_f \sim 2.6 N_c$. 
%When solving numerically for the running coupling we fix to the ultraviolet physics at some scale $\mu$ by setting $\alpha(\mu)$ fixed across theories with different $N_f$ (for example we choose $\alpha(e)=0.42$ for Fig. \ref{runingdm}). 

The one loop result for the anomalous dimension is
\begin{equation} \gamma = {3 C_2 \over 2\pi}\alpha= {3 (N_c^2-1) \over 4 N_c \pi} \alpha\,.  \end{equation}
So, using the fixed point value $\alpha_*$, the condition $\gamma=1$ occurs at $N_f^c \sim 4N_c$.

We will identify the RG scale $\mu$ with the AdS radial parameter $r = \sqrt{\rho^2+L^2}$ in our model. Note it is important that $L$ enters here. If it did not and the scalar mass was only a function of $\rho$ then were the mass to violate the BF bound at some $\rho$ it would leave the theory unstable however large $L$ grew. Including $L$ means that the creation of a non-zero but finite $L$ can remove the BF bound violation leading to a stable solution. Again in the D3/D7 system this mechanism is very natural as discussed in the Appendix.

\begin{figure}[]
\centering
%\hspace{-2mm}
\includegraphics[width=6.5cm]{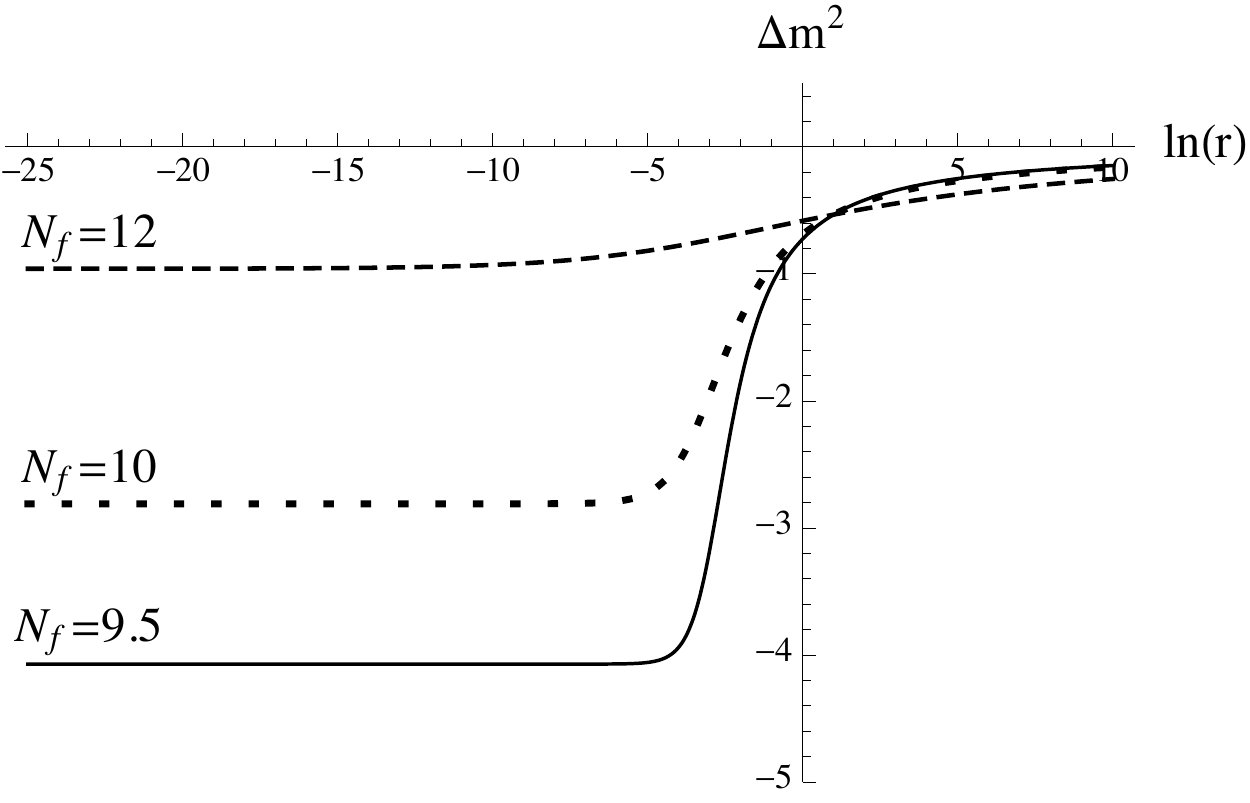}
\caption{$\Delta m^2$ vs $\ln(r)$for $N_c=3, N_f=12$, $10$, $9.5$ showing the increasing magnitude of the IR fixed point with lowering $N_f$.
Note that the BF bound is violated in the IR for $N_f <N_f^c \simeq12$. Here we are comparing theories with the same value of $\alpha$ at $\ln(r)=1$.}
\label{runingdm}
\end{figure}

Working perturbatively from the AdS result $m^2 = \Delta(\Delta-4)$ we have
\begin{equation} \label{dmsq3} \Delta m^2 = - 2 \gamma = -{3 (N_c^2-1) \over 2 N_c \pi} \alpha\, .\end{equation}
This will then fix the $r$ dependence of the scalar mass through $\Delta m^2$ as a function of $N_c$ and 
$N_f$. We sketch $\Delta m^2$ against $r$ for various cases in Fig. \ref{runingdm}, showing the IR fixed point behaviour. When solving numerically for the running coupling we fix to the ultraviolet physics at 
scale $\ln\mu=1$ by setting $\alpha(\mu)=0.1$ across theories with different $N_f$.

To completely specify the model we must also fix the parameter $\kappa$, which only enters into the computation of the axial-meson mass and decay constant. It is natural therefore to try to fix the parameter to give the observed mass splitting of vector and axial vector mesons in QCD. This leads us to fix $N_c=3$ for our analysis here and vary $N_f$. Of course we are really just choosing the RG flow of $\gamma$, and using perturbative results for this when $\alpha$ becomes large is at best questionable.  It is a model, and given this we will treat 
$N_f$ as a continuous parameter even at $N_c=3$ since then we will be able to smoothly move from a weakly coupled IR fixed point through to larger IR values and watch the behaviour of the spectrum. The generic features we will see should be applicable at other values of $N_c$. 

%In fact the simplistic modelling we have introduced only works when the running of the coupling is reasonable gentle and the final term in () is small. This means we can only compute down to $N_f=8.4$ or so. However, as we will see the spectrum saturates as one moves away from the chiral transition point and so fixing to the QCD splitting at $N_f=8.4$ seems reasonable. 

\begin{figure}[]
\centering
%\hspace{-2mm}
\includegraphics[width=6.5cm]{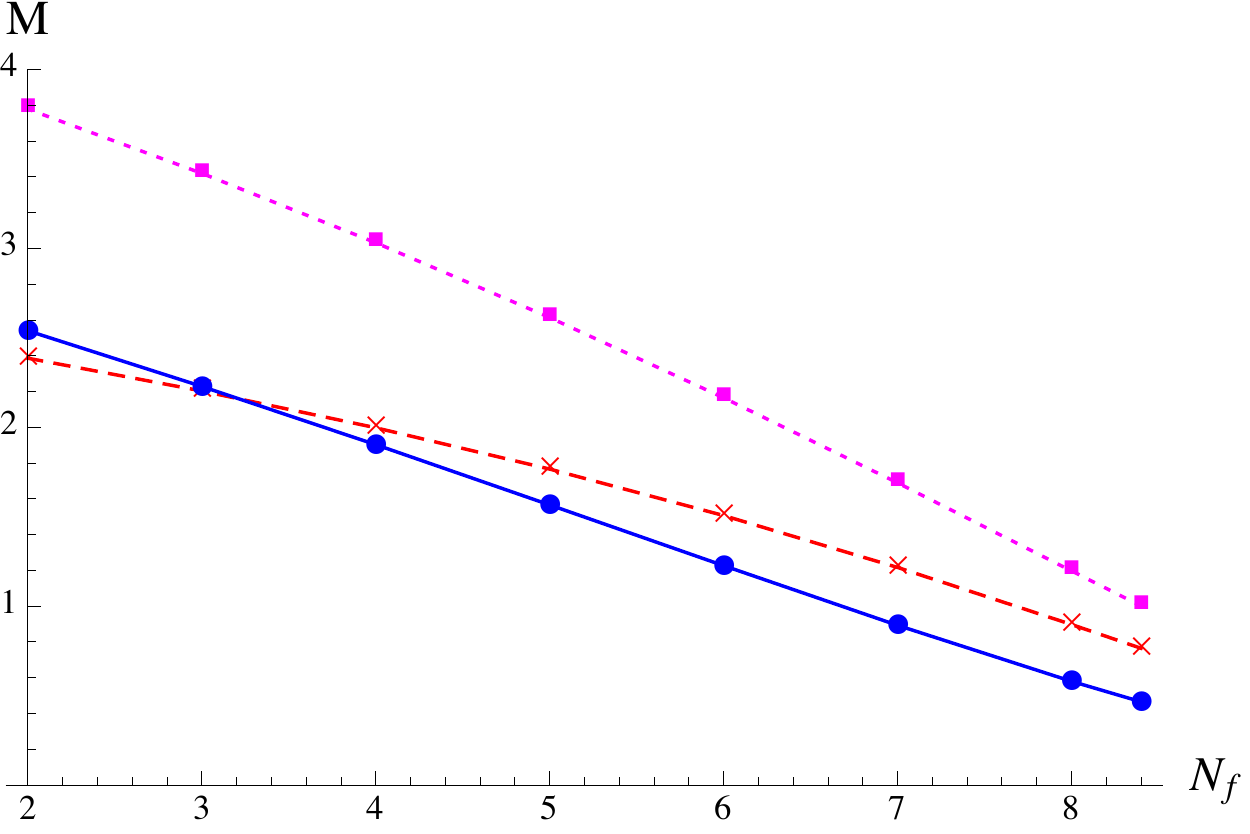}
\caption{Masses of the lightest scalar (dots), vector (crosses) and axial vector (squares) as a function of $N_f$. The values of $m_v/m_A$ at $N_f=2$ is input and the $N_f$ dependence is a prediction of the model.}
\label{massfull}
\end{figure}

We fix $\kappa$ so that the correct values of vector and axial-vector masses are reproduced at $N_f=2$. Then we must further choose how to make $\kappa$ scale with $N_f$. As discussed above we expect axial-vector symmetry to be smoothly restored at the continuous chiral phase transition. We therefore require $\kappa \rightarrow 0$ 
at the transition at $N_f^c (\simeq 12)$. Taking all of these considerations into effect we choose
\begin{equation} \label{k3} \kappa^2 = 3.34 (N_f- N_f^c). \end{equation}
Although reasonably motivated, this choice is somewhat adhoc. Again, though, we expect the broad behaviours of the spectrum to be correctly modelled; the results for the masses of the scalar, vector and axial mesons are shown in Fig. \ref{massfull} as a function of $N_f$. The value of $m_V/m_A$ at $N_f=2$ is input to constrain the model while the $N_f$ dependence is a prediction. Our main interest is in the large-$N_f$ region and these results will be discussed in more detail below.

\subsection{Results at $m_q=0$}

\begin{figure}[] 
\centering
%\hspace{-2mm}
\includegraphics[width=6.5cm]{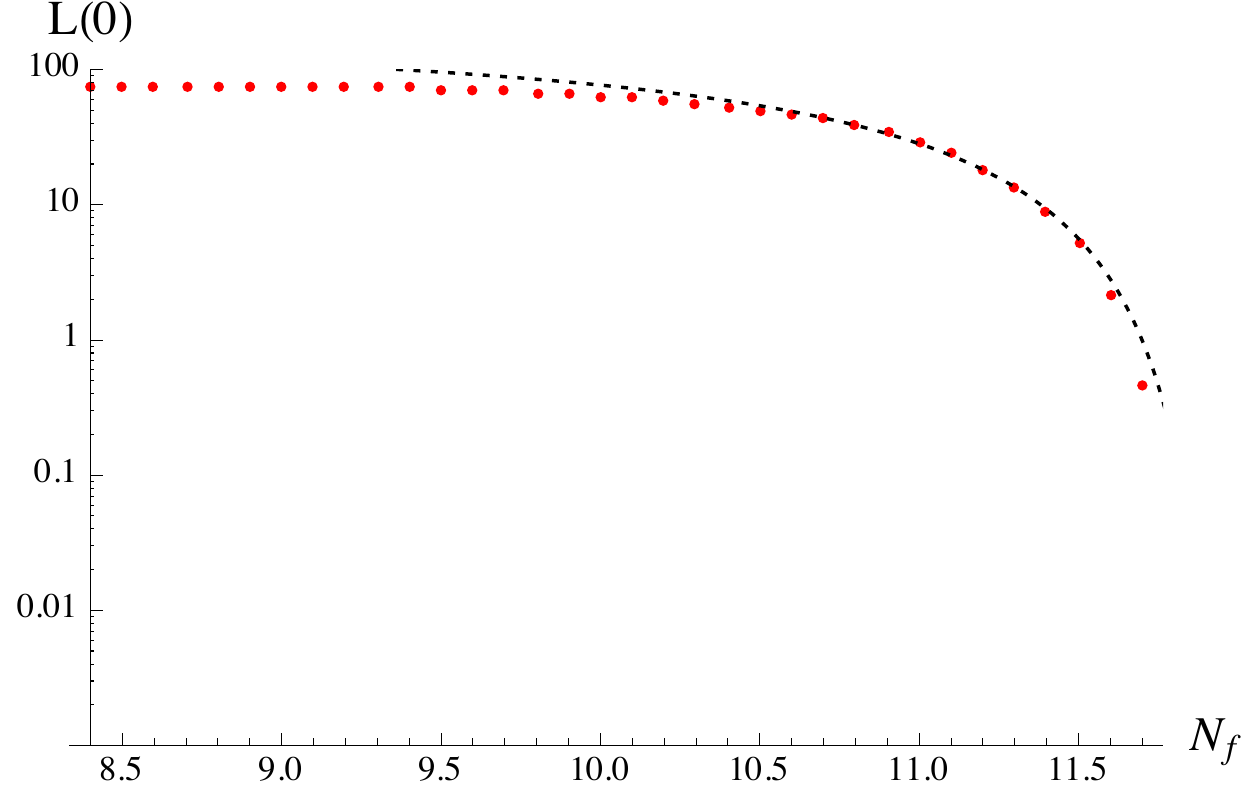}
\caption{The IR quark mass, L(0) vs $N_f$. The dashed line shows the fit $a \exp[-b/(12-N_f)^{1/2}] (N_f^c-N_f)^{p_L}$, with $a=4692.42$, $b=5.111$ and $p_L=0.722$.}
\label{lplot}
\end{figure}

The model is now completely fixed (there are no free parameters) and we can compute in the zero quark mass limit. First we solve Eq. (\ref{embedeqn}) for the profile of the field $L$. The value of $L(0)$ is a measure of the dynamical quark mass in the IR and we plot it against $N_f$ in Fig \ref{lplot}. 
%It shows clear BKT scaling as we approach $N_f^c$. 
Note that for $N_f < 9.5$ the value of $L(0)$ saturates to a near constant. This seems reasonable because for all these theories the scale at which the BF bound violation occurs is very similar, not separated by orders of magnitude of running (because of our initial condition on $\alpha$). One would expect very similar IR physics. Above $N_f = 9.5$ the theories still have considerable running time between the fixed UV scale and the IR scale at which the BF bound is violated. For this reason $L(0)$ falls and shows considerable $N_f$ dependence. The scaling is of BKT type modulated by a power law in $N_f^c-N_f$,
\beq
L(0)=a \exp\left(-b/\sqrt{N_f^c-N_f}\right) (N_f^c-N_f)^{p_L},
\eeq
where $a=  4692.42$, $b=-5.111$ and $p_L=-0.722$. Numerical solution for $L(0)$ and the above fit are shown in Fig \ref{lplot}. Hence, we expect that the transition to the conformal window from the chiral symmetry breaking phase is continuous and displays Miransky scaling. 
Of course all dimensionful physical quantities in the theory are expected to show similar scaling. In practice, when fitted over a finite range in $N_f$ the scaling behaviour of all quantities can be characterised by a BKT type behaviour, but with different fit parameters for different quantities. This suggests that while there is an IR scale which shows the expected BKT behaviour and to which all other physical quantities are proportional, there can be additional $N_f$-dependent factors enhancing the scaling. This is precisely the behaviour we have seen for $L(0)$ above. 
%This is precisely what we find: the IR observable which scales towards zero exponentially fast is $L(0)$. 
%All masses and decay constants are proportional to $L(0)$, but with additional suppression factor $(N_f-N_f^c)^p$.

\begin{figure}[]
\centering
%\hspace{-2mm}
\includegraphics[width=6.5cm]{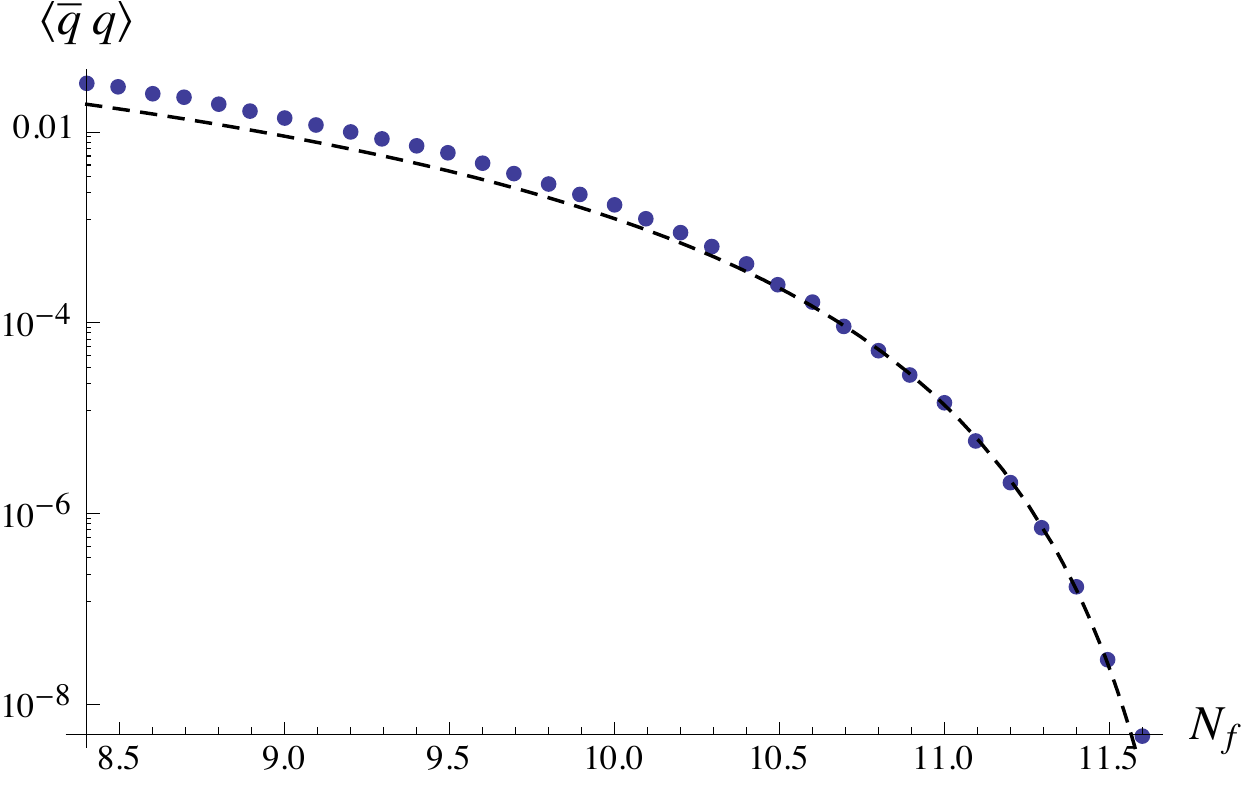}
\caption{The dots show numerical results for the quark condensate as a function of $N_f$. The dashed line is the BKT fit  $a \exp(-3 b/(N_f^c-N_f)^{1/2})$ with parameters $a=63.090$ and $b=5.111$.}
\label{qbarq}
\end{figure}

It turns out that there is a quantity, the quark condensate, which displays BKT scaling, i.e. a pure exponential form $a \exp(-3 b/(N_f^c-N_f)^{1/2})$.  
In our model, we can determine the quark condensate from the asymptotic UV behaviour of the profile $L(\rho)$ which is in the $m_q=0$ limit given by
\beq \label{con}
L(\rho)=\langle\bar{q}q\rangle \rho^{-2} \left(\ln\rho\right)^{k},
\eeq 
where $k=3 C_2/(4\pi b_0)$. The logarithmic correction is due to the 
running of the coupling. 
%The dimension 3 UV quark condensate is proportional to the coefficient of this term.
We show the numerical results  for the condensate as well as the fit of the BKT form with parameters $a=63.090$ and $b=5.111$ in Fig. \ref{qbarq}. 

\begin{figure}[]
\centering
%\hspace{-2mm}
\includegraphics[width=6.5cm]{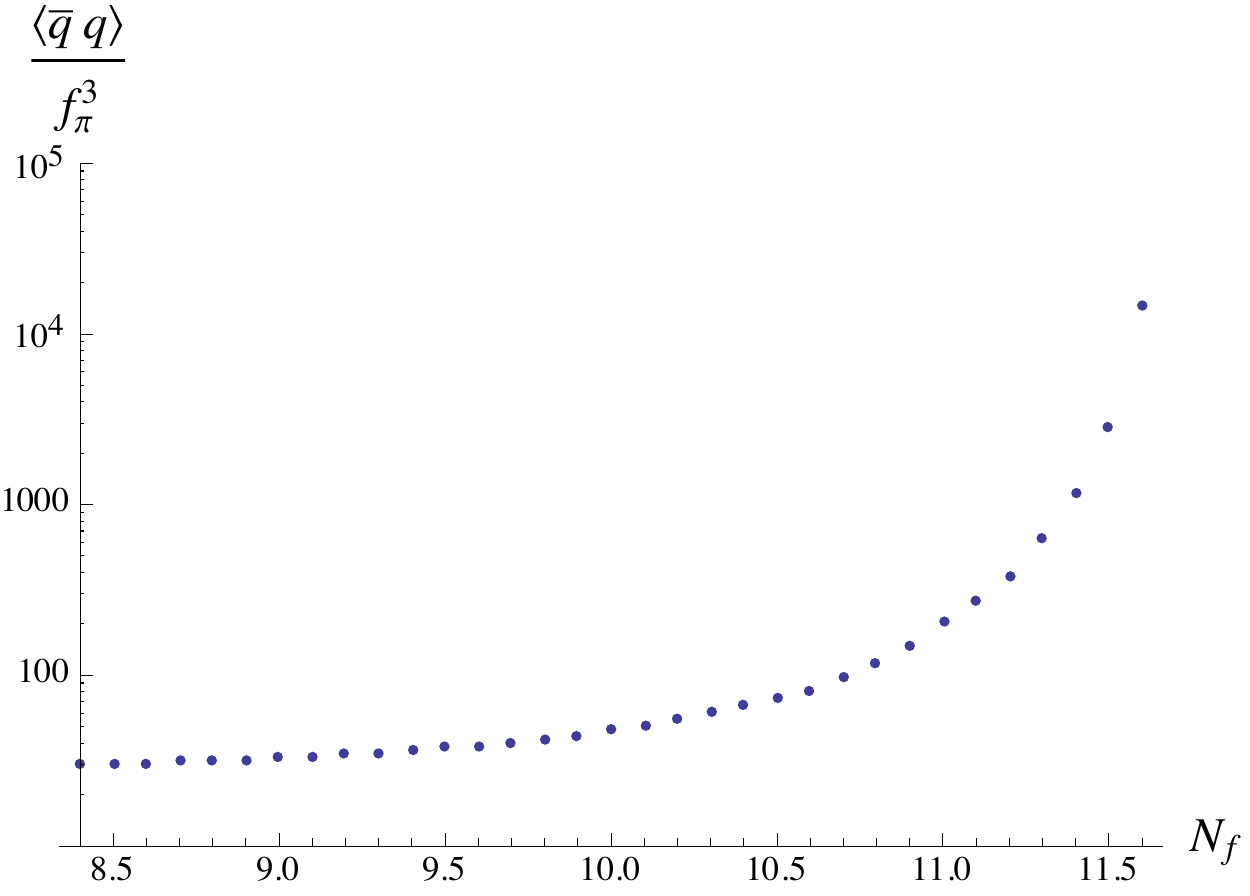}
\caption{The quark condensate normalized by $f_\pi^3$ vs $N_f$. }
\label{cond}
\end{figure}

The classic expectation of walking gauge theories is that theories, that have a large range over which 
$\gamma \simeq 1$, will have an enhanced UV quark condensate (the UV dimension 3 condensate is expected to be given by the product of the IR dimension 2 condensate and the UV scale at which 
$\gamma$ transitions to its higher value). 

We plot the dimensionless $\langle\bar{q}q\rangle/f_\pi^3$ against $N_f$ in Fig. \ref{cond}. 
When $N_f$ is sufficiently far below the conformal window, our results show the expected 
$\langle\bar{q}q\rangle\sim f_\pi^3$. However, when $N_f$ is increased, the condensate is enhanced relative to 
$f_\pi^3$. This enhancement of the condensate in the walking regime near the phase transition is a very clear prediction of the model.

\begin{figure}[] 
\centering
%\hspace{-2mm}
\includegraphics[width=6.5cm]{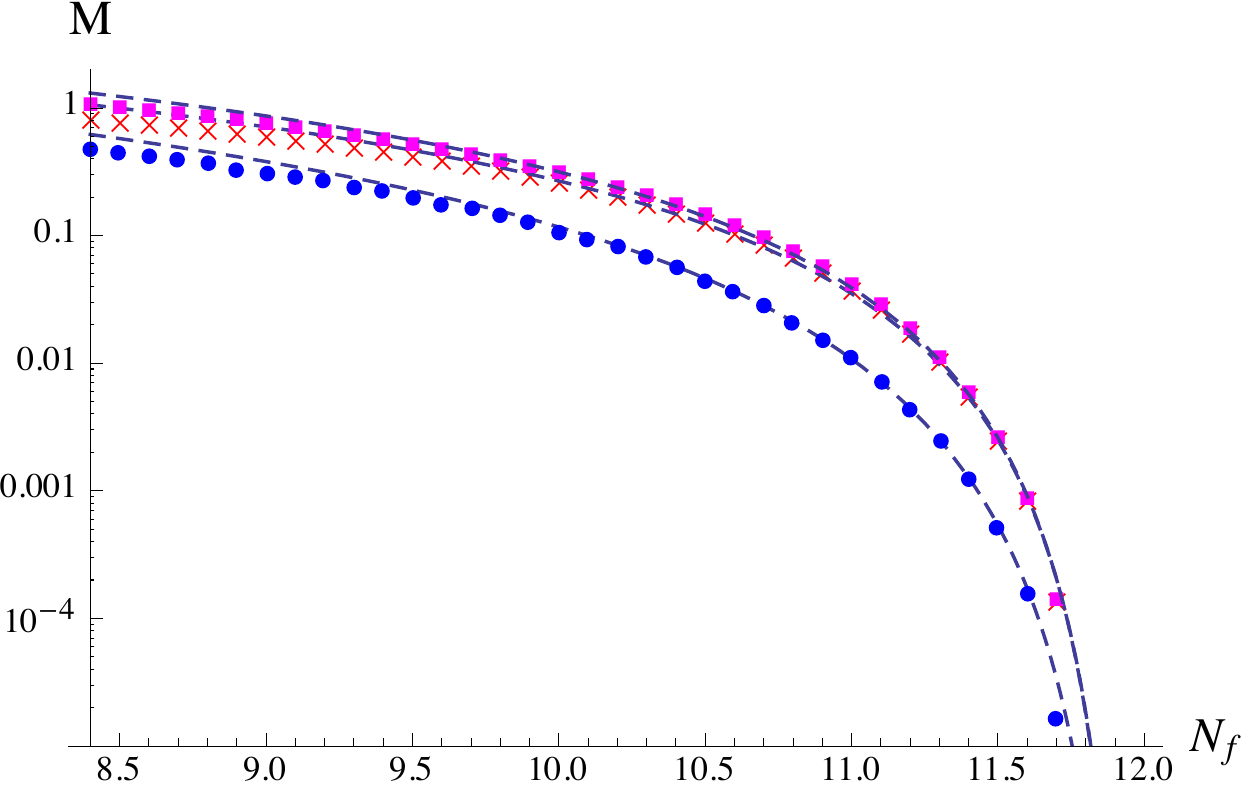}
\caption{Masses of scalar (dots), vector (crosses) and axial (squares) mesons as a function of $N_f$. 
The dashed lines show fits of the form $M_i=(N_f^c-N_f)^{p_{i}}L(0)$, where $p_S= 1.27,$ 
$p_V=0.77$ and $p_A=0.85$.}
\label{svafitplot}
\end{figure}

We can now turn to the bound state masses and decay constants.
As we have already discussed for the case of $L(0)$, we expect that the scaling of all physical scales near $N_f^c$ is  
\beq
O_i=A_{i}(N_f^c-N_f)^{p_i} \langle\bar{q}q\rangle^{1/3},
\label{massfit}
\eeq
where the overall normalization $A_i$ and the power $p_{i}$ are constants dependent only on the observable in question. Let us show this in detail for the masses of the scalar ($S$), vector ($V$) and axial ($A$) mesons. We find that each of these quantities falls to zero according to Eq. (\ref{massfit}) as shown in Fig. \ref{svafitplot}. In the figure the numerical results for the scalar masses are shown by dots, vector masses by crosses and axial meson masses by squares. The dashed lines are the fits of the form (\ref{massfit}) with $p_{S}=1.27$, $p_{V}=0.77$ and $p_{A}=0.85$. 
The fit for the pion decay constant is also of the same form and gives $p_{f_\pi}=1.02$. 

Hence, we find that the masses scale towards zero faster than the pure BKT behaviour given by the behaviour of $\langle \bar{q}q\rangle$ alone. Among the masses the scalar mass behaves differently than the vector and axial meson masses. These results illustrate that one must be careful when extracting the scaling behaviours of various quantities with $N_f$ from lattice simulation results where only a finite range of discrete values of $N_f$ is accessible.

\begin{figure}[]
\centering
%\hspace{-2mm}
\includegraphics[width=6.5cm]{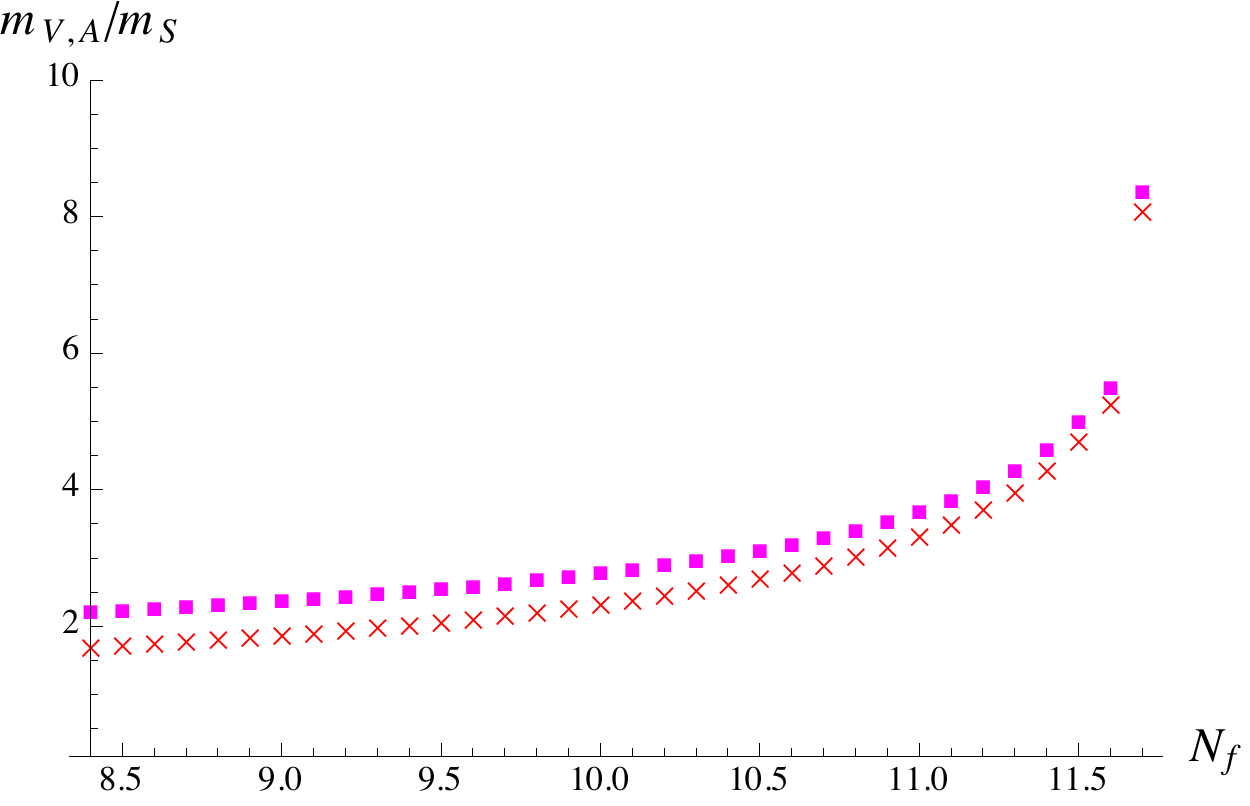} \\  
\includegraphics[width=6.5cm]{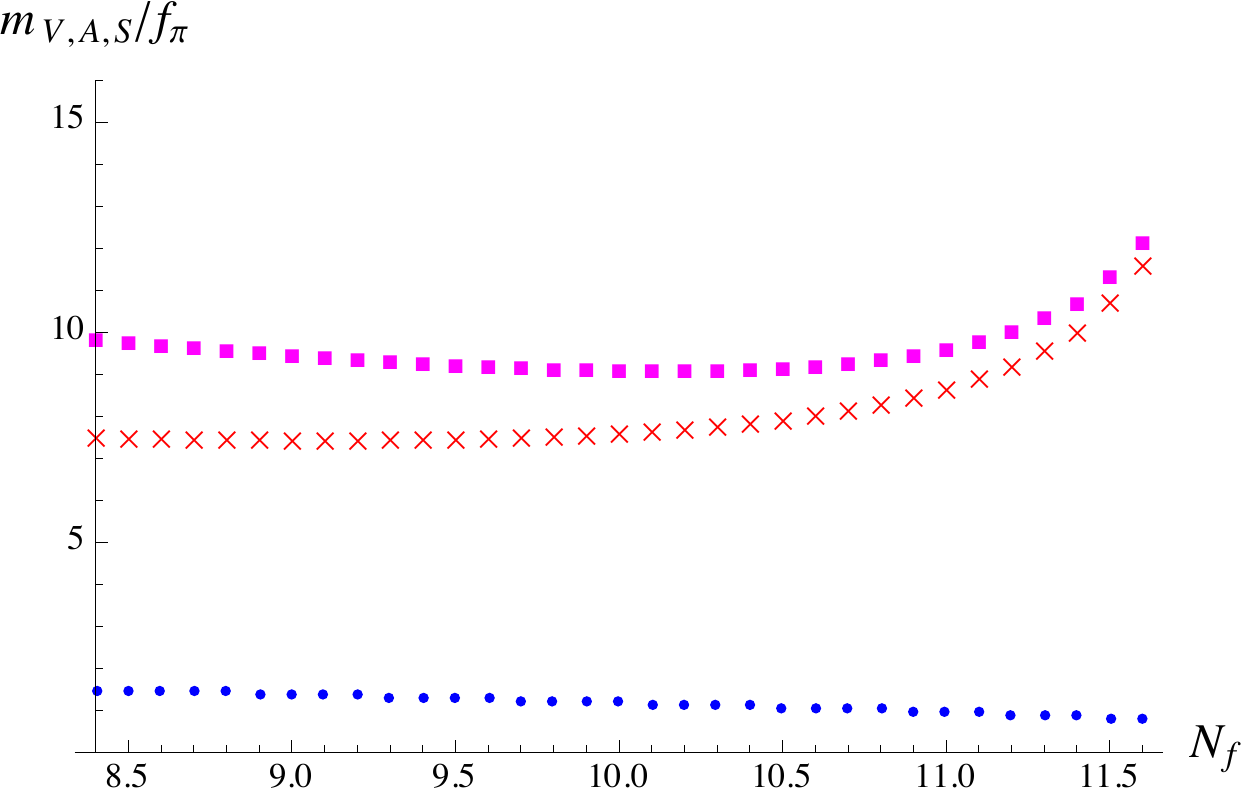}
\caption{Mass ratios: in the top figure the crosses and squares show the ratio of the $V$ and $A$ masses to the $S$ mass, respectively, revealing the Goldstone-like nature of the scalar meson as one approaches the critical point. The bottom figure shows the masses in units of $f_\pi$. The squares denote $m_A$, crosses $m_V$ and dots $m_S$.}
\label{mr}
\end{figure}

One of the phenomenologically most interesting questions is the ratio of these masses. 
We plot the ratio of the $V$ and $A$ masses to the scalar ($S$) mass in the top plot of Fig. \ref{mr}. It is clear that the scalar mass becomes light as one approaches the chiral symmetry transition at $N_f^c\simeq 12$. As in \cite{Evans:2013vca} our interpretation is that the potential for $\langle\bar{q}q\rangle$ becomes very flat as one approaches the transition - the large anomalous dimension of $\langle\bar{q} q\rangle$ over a wide running range leads to the enhancement of its vev as we have seen. The potential difference between the vacuum value and 
$\langle\bar{q} q\rangle=0$ is controlled by the IR, though, so it is order $f_\pi$. Thus as one approaches the transition the potential becomes arbitrarily flat and a Goldstone associated with shift symmetry in energy emerges - the light $S$ meson. Note that in \cite{Evans:2013vca} we showed that the scalar became light relative to $L(0)$ but here in this extended model we can see its Goldstone like behaviour relative to the 
$V$ and $A$ masses, $f_\pi$ and also the other decay constants as we will see below. 
The lightness of this state near the transition is rather clear in this model.

The lower plot in Fig. \ref{mr} shows the masses of the three mesons in units of $f_\pi$. As the transition point is approached the $V$ and $A$ become degenerate by construction through our choice of the $N_f$ dependence in the parameter $\kappa$. At $N_f=2$ the ratio of the $V$ and $A$ masses matches that in QCD again by assumption. 

\begin{figure}[]
\centering
%\hspace{-2mm}
\includegraphics[width=6.5cm]{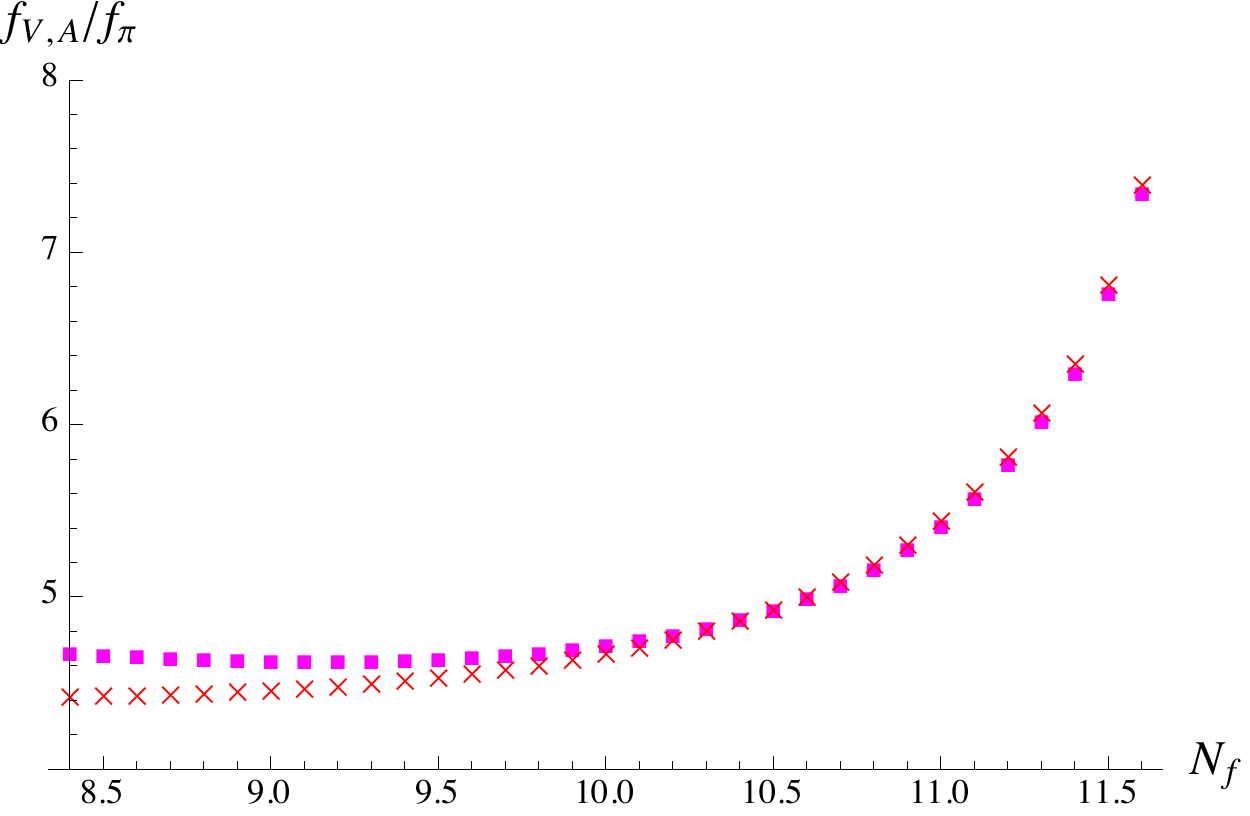} 
\caption{The $V$ and $A$ decay constants, shown by crosses and squares respectively, in units of $f_\pi$ against $N_f$.}
\label{dcVA}
\end{figure}

\begin{figure}[]
\centering
%\hspace{-2mm}
\includegraphics[width=6.5cm]{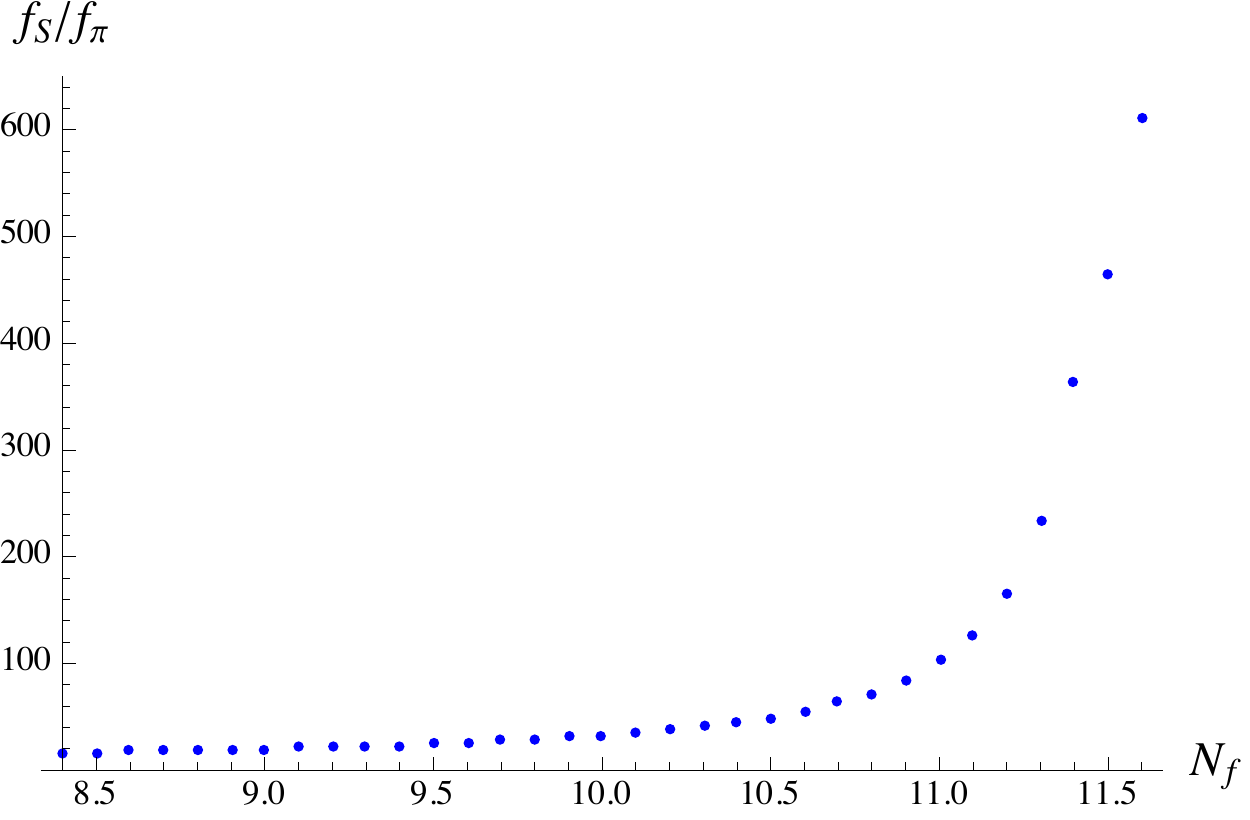} 
\caption{Decay constant of the scalar meson divided by $f_\pi$ as a function of $N_f$.}
\label{dcS}
\end{figure}

The decay constants for the $V$ and $A$ in units of $f_\pi$ are shown in Fig. \ref{dcVA}. Again by construction they become degenerate near the transition. In Fig. \ref{dcS} we show the decay constant of the scalar meson divided by $f_\pi$. For $N_f$ sufficiently far below the conformal window we observe that the
obvious expectation of a QCD-like theory is satisfied as $f_S\sim f_\pi$. However, as $N_f$ is increased towards to boundary of the conformal window we see a large hierarchy between $f_S$ and $f_\pi$ arising.
This constitutes yet another clear prediction of our model in the walking region. 

To conclude this analysis, we discuss the implications on electroweak physics. 
If one were to imagine using the gauge theories of the type we have analyzed here as 
technicolour models of electroweak symmetry breaking, then a key quantity to confront with existing 
electroweak data are the oblique corrections, in particular the S parameter \cite{Peskin:1991sw}. The S parameter is given by
\begin{equation} S = 4 \pi \left( \Pi_{AA}'(0) - \Pi_{VV}'(0) \right)\,, \end{equation}
where $\Pi$ are the vector vector and axial axial correlators. The derivative is with respect to $q^2$ and the derivatives are evaluated at $q^2=0$. 

In the UV the correlators are given by (\ref{Ks}) and since the axial symmetry is unbroken the contribution to S is zero. Our insistence on the restoration of the vector-axial symmetry in the spectrum as we approach $N_f^c$ will also mean that the S parameter vanishes at the BKT transition point. 

S is usually computed in technicolor models through the contributions of the various bound states of the theory. 
For a model to be realistic in modelling the electroweak symmetry breaking of the Standard Model, the 
scalar meson must mimic the Standard Model Higgs with mass 125 GeV (and appropriate couplings). Its loop contributions will, we assume, therefore be that of the SM.

\begin{figure}[]
\centering
%\hspace{-2mm}
\includegraphics[width=6.5cm]{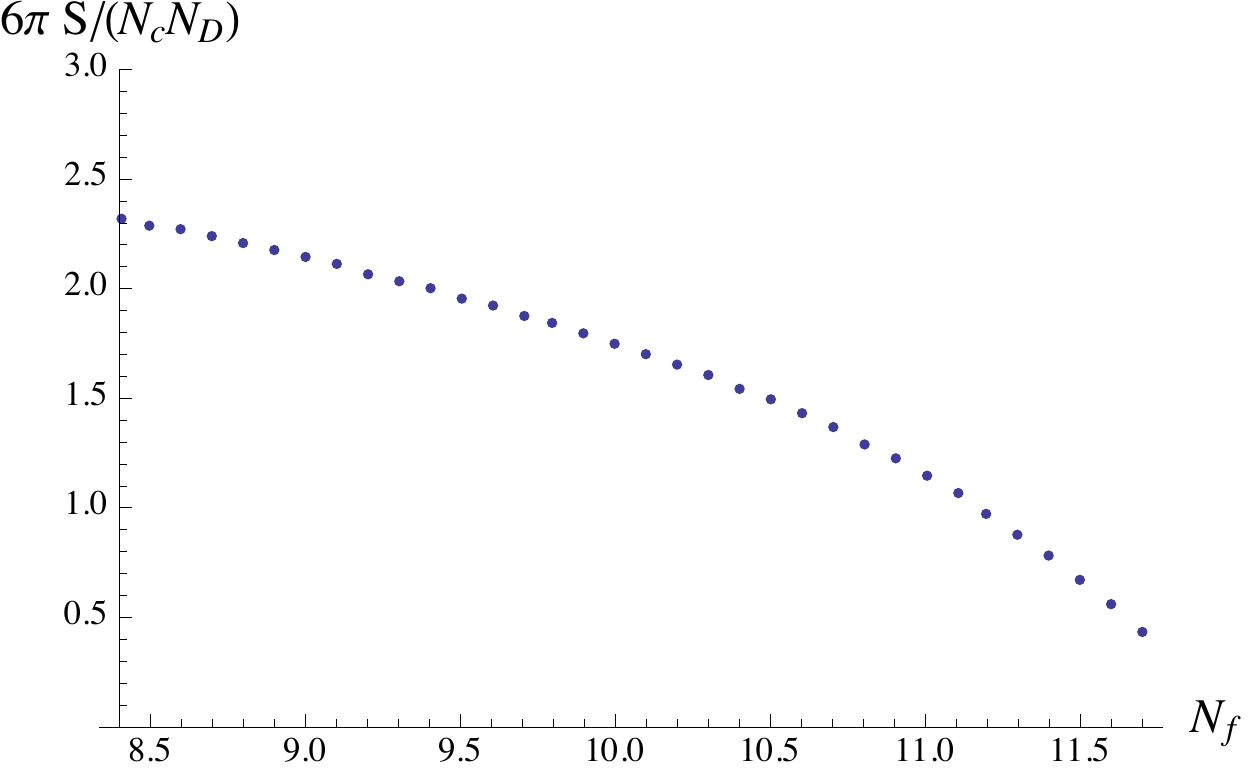} 
\caption{The contribution to the S parameter (normalized by the number of techni-doublets and to the perturbative value from a single mass degenerate doublet) from the lightest vector and axial mesons.}
\label{s}
\end{figure}

The  $V$ and $A$ mesons will make a further contribution which is of size
\begin{equation}
S = 4 \pi  \left({f_V^2 \over m_V^2} -  {f_A^2 \over m_A^2}\right)\,.
\end{equation}
We plot this extra contribution, normalized by the number of electroweak doublets,  in Fig \ref{s}. The behaviour sensibly matches expectations: for QCD-like theories at lower $N_f$ the S contribution per doublet is larger (by an order one number) than the contribution of a mass degenerate perturbative doublet ($1/6 \pi$). In the walking regime the contribution per doublet falls to zero at the chiral transition point.

\subsection{Results at $m_q\neq 0$}

Finally let us investigate the effects of small finite quark mass in the BKT region. In the massive case the UV profile $L(\rho)$ has an additional non-normalizable piece (on top of the normalizable mode in (\ref{con}))
\begin{equation} L(\rho) = {m \over ({\rm ln} \rho)^k} \end{equation} with $m$ interpreted as the quark mass. Note the dimension of the product $m \bar{q}q$ is always 4 in the model. 

\begin{figure}[]
\centering
%\hspace{-2mm}
\includegraphics[width=6.5cm]{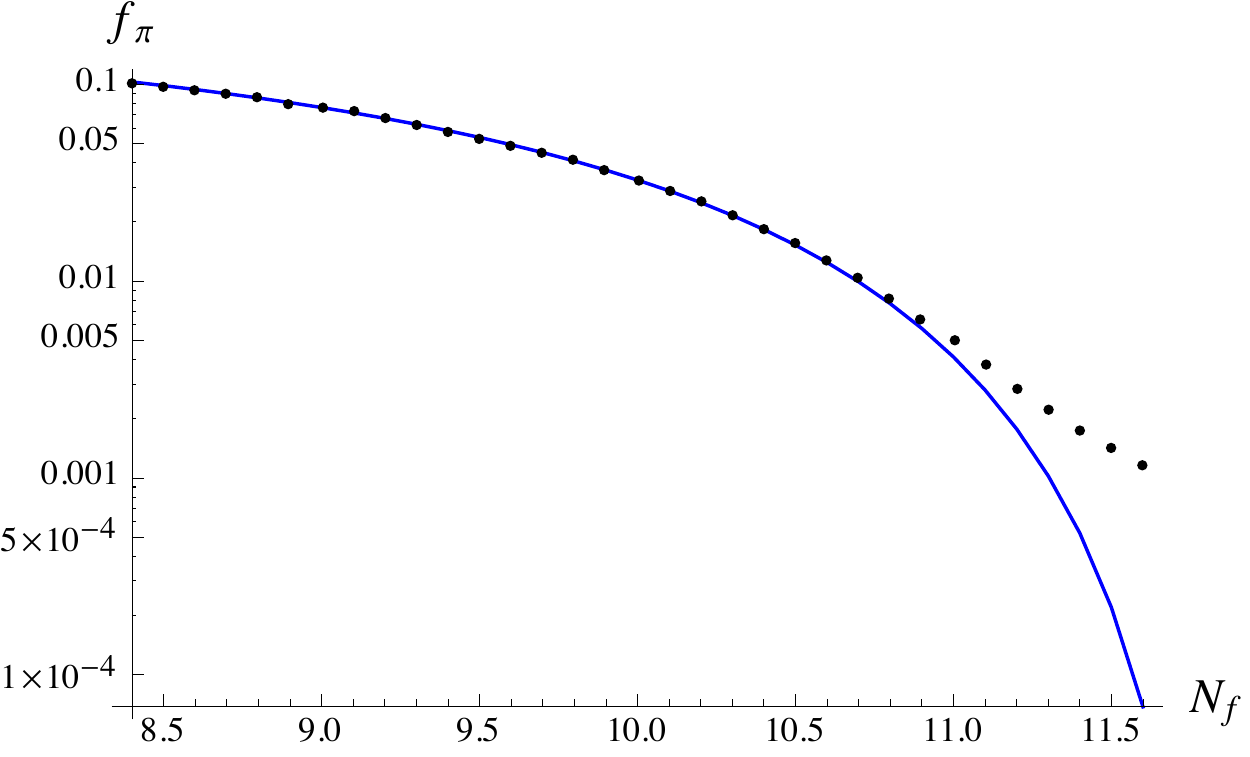} 
\caption{The pion decay constant plotted against $N_f$. Solid line corresponds to $m_q=0$, while the dots correspond to $m_q=10^{-5}$.}
\label{fpimq}
\end{figure}

In Fig. \ref{fpimq} we show the pion decay constant for the cases $m_q=0$ and $m_q=10^{-5}$. We see that, while the finite quark mass curve traces well the corresponding zero mass curve at low $N_f$, it deviates as $N_f^c\simeq 12$ is approached.
The transition corresponds to the point where the IR scale where the BF bound is violated becomes equal to the hard quark mass. Note that the bound states persist above $N_f^c$ since the hard mass breaks conformality, and there is no conformal window. 

%\begin{figure}[]
%\centering
%%\hspace{-2mm}
%\includegraphics[width=6.5cm]{L0BFKBreaking.pdf} 
%%
%\caption{The IR quark mass $L(0)$ plotted agnst $N_f$  Red: $mq=0$, Blue: $m_q=10^{-5}$.}
%\label{L0mq}
%\end{figure}

\begin{figure}[]
\centering
%\hspace{-2mm}
\includegraphics[width=6.5cm]{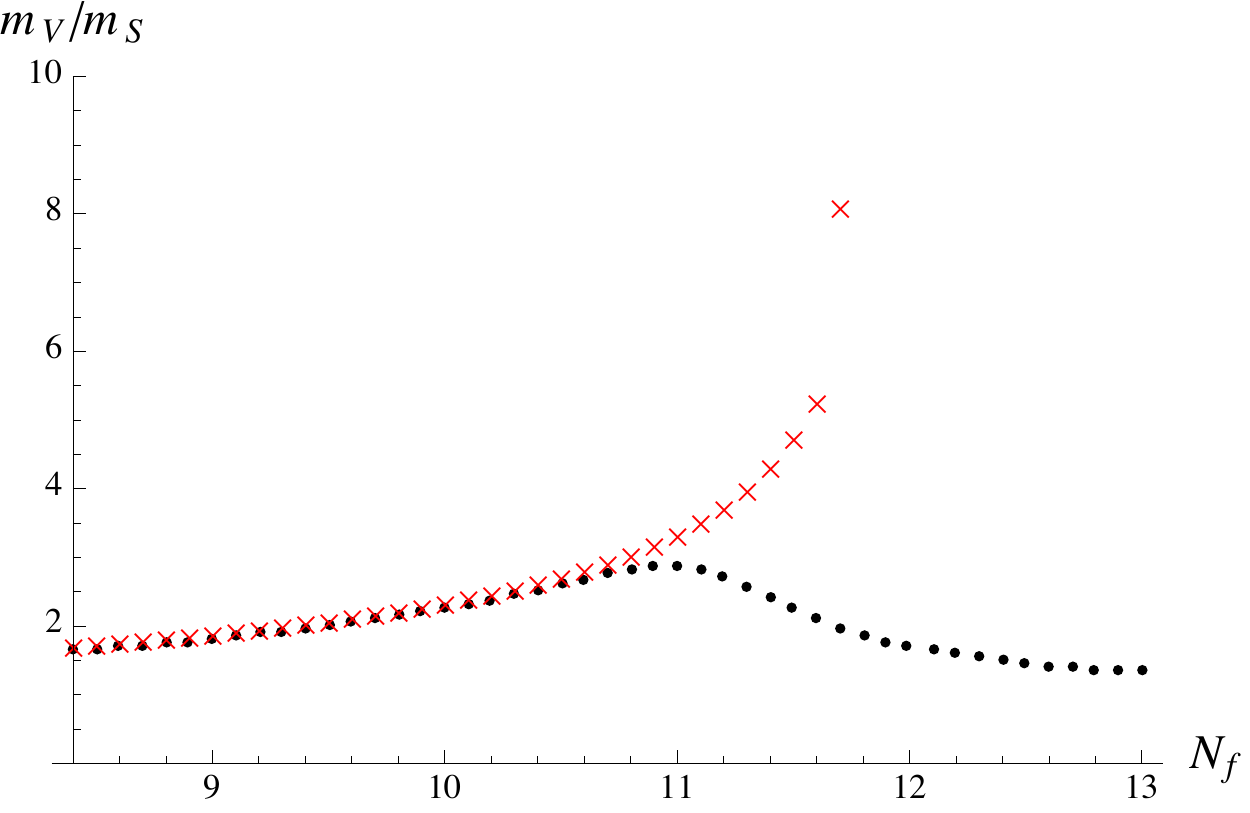} 
\caption{The ratio of $m_V/m_S$ against $N_f$. The crosses show the vector meson mass in the $mq=0$ case while the black dots correspond to $m_q=10^{-5}$.}
\label{spervmq}
\end{figure}

The Goldstone nature of the scalar is also lost; Fig. \ref{spervmq} shows $m_V/ m_S$ against $N_f$. Again, below $N_f=12$ the finite mass behaviour coincides with the zero mass case, but as $N_f\simeq 12$ is approached all states begin to scale with $m_q$ leading to the degeneracy implied by the figure.

\begin{figure}[]
\centering
%\hspace{-2mm}
\includegraphics[width=6.5cm]{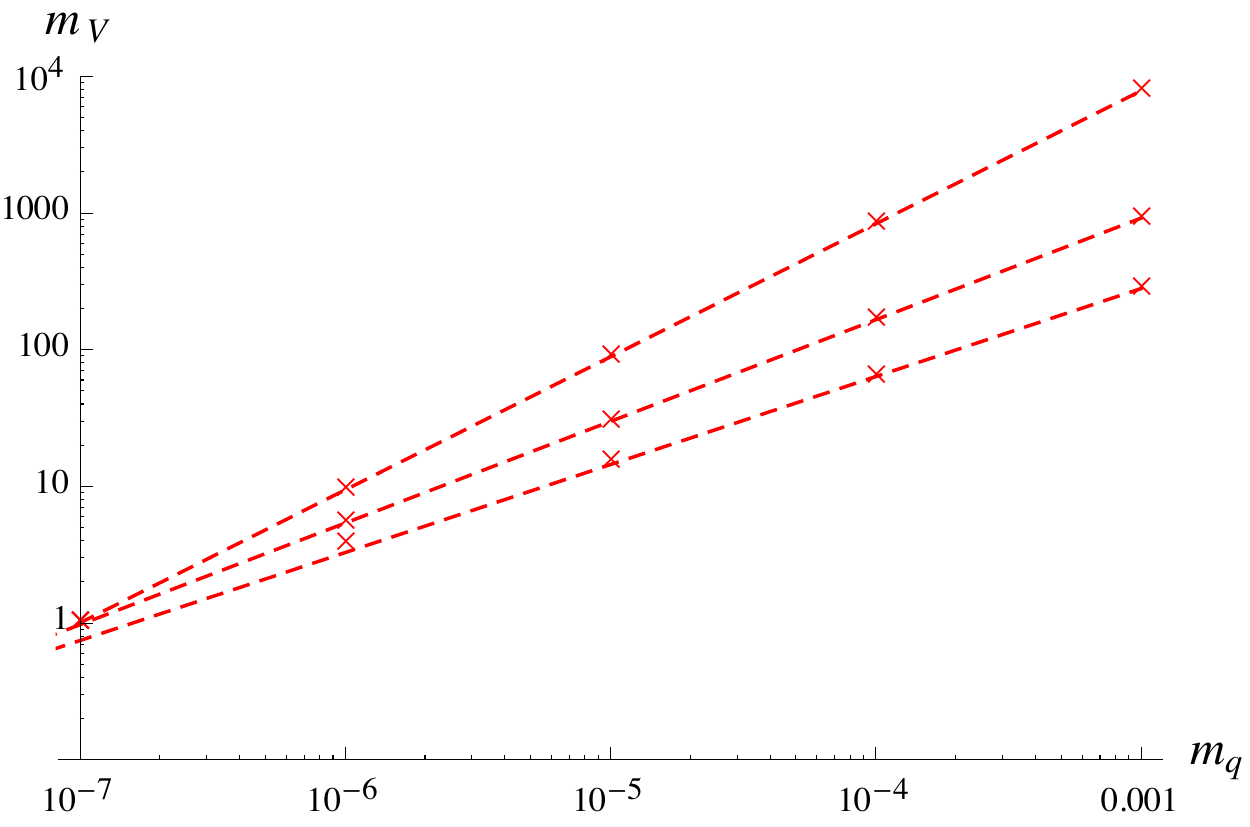} 
\caption{$m_V$ vs $m_q$ at $N_f=16,13,12$ from top to bottom. The dotted lines are fits of the form $m_q^b$ with $b=0.644$ for $N_f=12$, $b=0.744$ for $N_f=13$ and $b=0.974$ for $N_f=16$. }
\label{yes}
\end{figure}

To test the scaling behaviour of the masses we plot the $V$ mass against the current quark mass at $N_f=12,13,16$ in Fig. \ref{yes}. They are well fitted by the curves $m_q^b$ with $b=0.644$ $(N_f=12)$, $b=0.744$ $N_f=13$ and $b=0.974$ $N_f=16$. These powers are readily explained. For small enough quark mass, the quark mass has dimension $1 + \gamma_*$, with $\gamma_*$ the IR fixed point value in the conformal regime. Hence the dimension one $\rho$ mass is expected to scale as $m_q^{1/(1 + \gamma_{*})}$ \cite{DelDebbio:2010ze} - this predicts $b=0.676, N_f=12$, $b=0.770, N_f=13$ and $b=0.974, N_f=16$. The agreement is satisfyingly at the few percent level.

\section{Discussion}

We have introduced a variant of AdS/QCD which has a dynamical mechanism for the generation of the quark condensate and naturally introduces a soft wall at the scale of the quark condensate. The model is basically just the linearized form of the top-down D3/D7 model but with the running of the anomalous dimension of the quark condensate input by an ansatz through the AdS scalar mass. We have used the model to study the dynamics of SU(3) gauge theory with $N_f$ quarks (we keep $N_f$ as a continuous parameter in the parametrization of $\gamma$). The model is then completely fixed by its action (\ref{daq}), the imposition of the radially dependent mass using perturbative QCD results in (\ref{dmsq3}) and the choice of the coupling $\kappa$ in (\ref{k3}). The latter is picked to match the ratio of vector and axial vector meson masses in $N_f=2$ QCD and to scale with $N_f$ so that the axial vector symmetry is restored at the continuous chiral transition at the edge of the conformal window. 

We find it remarkable that such a basic model then rather simply reproduces the entire lore about walking technicolor dynamics. Chiral symmetry breaking is triggered by the anomalous dimension of the quark bilinear, $\gamma$, growing above 1. The transition displays Miransky scaling (Fig. \ref{lplot}) at the chiral restoration transition.  In the walking regime the quark condensate grows relative to $f_\pi$ (Fig. \ref{cond}), the scalar meson becomes light relative to the rest of the spectrum (Fig. \ref{mr}), and the electroweak S parameter falls to zero  (Fig. \ref{s}). 

When a small current quark mass is introduced into the theory the chiral transition is lost and the spectrum moves at large $N_f$ to a scaling behaviour where all dimensionful quantities scale as the appropriate power of $m_q$ (given its anomalous dimension at the IR fixed point). This provides an explicit model realizing the scalings proposed in \cite{DelDebbio:2010ze} and should provide a very helpful guide for lattice simulations of these theories.

The model, of course, is simplistic, not least in using the perturbative running of the gauge coupling. For example recent lattice simulations for SU(3) with $N_f=12$ have predicted IR fixed point values for $\gamma$ in the range of 0.386 \cite{Appelquist:2011dp}, 0.459 \cite{Aoki:2012eq}, and 0.32 \cite{Cheng:2013eu} to be compared with the one loop perturbative value 0.48 we use. Although the precise runnings are open to correction the broad flavour of our results is likely to be correct assuming the fixed point behaviour persists to sufficiently high values of $\gamma$ (the possibility of ``jumping" \cite{Sannino:2012wy} exists and would give very different results).

Finally, it remains a fascinating question as to whether such physics could still be compatible with LHC data.
Technicolor dynamics remains very appealing because it removes fundamental scalars from the standard model and allows the physics of flavour to live at  scales that can potentially be probed experimentally.  Walking does seem capable of hiding many of the deficiencies of technicolor dynamics. For example, our model suggests a rather rapid growth of the quark condensate if $N_f$ lies within 10$\%$ of its critical value. This would alleviate problems with flavour changing neutral currents and the T parameter \cite{Chivukula:1995dc} from extended technicolor. In the same regime the scalar meson becomes a light Higgs like state with mass of order $f_\pi$ with as much as an order of magnitude gap between its mass and those of other bound states. In this regime the S parameter contribution of a doublet is also smaller than a perturbative doublet leaving more room in precision data. The tuning needed to achieve these results seem to be alleviated in these models because tuning to $N_f^c$ simultaneously tunes $\gamma \rightarrow 1$ and increases the running distance over which near conformality is present. Both features are phenomenologically helpful. A key challenge remains the rate of the scalar meson decay to two photons which we have not addressed here. Of course, even if nature does not use these schemes, the study of the conformal window remains an important theoretical question and we hope this model will be useful in motivating lattice simulations of these systems. 

\bigskip \bigskip

\noindent {\bf Acknowledgements:}  NE is grateful for the support of a STFC consolidated grant. TA
thanks the V\"ais\"al\"a foundation for financial support.
KT acknowledges support from the European Science Foundation (ESF) within the framework of the ESF 
activity entitled `Holographic Methods for Strongly Coupled Systems', and hospitality at University of 
Southampton during the time the research reported here was carried out.\newpage

\section{Appendix: The D3/probe D7 system}

The dynamics of the condensate formation and how it forms a soft wall in meson computations in Dynamic AdS/QCD is based on the D3/probe 
D7 model \cite{Karch:2002sh}. The D3 branes generate the duality between ${\cal N}=4$ super Yang-Mills theory and the space AdS$_5\times$S$^5$,
 \begin{eqnarray}
ds^2 &=& G_{MN} dx^M dx^N \nonumber \\
&&\\
&=& {r^2 \over R^2} dx_{3+1}^2 + {R^2 \over r^2} ( d \rho^2 + \rho^3 d \Omega_3^2 + dL^2 + L^2 d\phi^2),
\nonumber \end{eqnarray} 
where $R$ is the AdS radius and $r^2=\rho^2 + L^2$. To introduce ${\cal N}=2$ quark hypermultiplets into the gauge theory, probe D7 branes can be introduced in the $x_{3+1}, \rho, \Omega_3$ directions ($\xi^a$). They then lie at fixed $\phi$ with a potentially non-trivial profile $L(\rho)$.  That profile is determined by the DBI probe action (in Einstein frame)
\begin{equation}
S_7 = T_7 \int d^8\xi e^{\phi} \sqrt{P[G_{ab}]}, 
\end{equation}
where $\phi$ is the dilaton (dual to the gauge coupling). The pulled back metric seen by fields on the D7 world-volume is
\begin{eqnarray}
 P[G_{ab}] & = & G_{MN} {dx^M \over d \xi^a} {dx^N \over d \xi^b}. 
 \end{eqnarray}
We have
\begin{equation} \label{this} S_7 = T_7 \int d\rho e^{\phi} \rho^3 \sqrt{1 + (\partial_\rho L)^2 +  {R^4 \over r^4 }(\partial_x L)^2 }. \end{equation}

In the supersymmetric case the dilaton is constant and the linearized equation of motion for the vacuum configuration $L(\rho)$ is just the first term in (\ref{embedeqn}) - as is well known the embedding profile encodes the quark mass and condensate holographically. The equation for the mesonic fluctuations about the vacuum embedding are then the first and last term in (\ref{deleom}). Note that it is crucial that the action has a pre-factor of $\rho^3$ (not $r^3$) for the correct holographic relations to emerge. It is also crucial that in the embedding equation a factor of $r^2 = \rho^2 + L^2$ is present cutting off the space at $r$ of order the IR quark mass - if it were not there and the computation extended to $r=0$ the spectrum would become conformal. We have carried both of these aspects of the model across into our Dynamic AdS/QCD model. 

To understand how a radially dependent mass term for $L$ can emerge we can consider a case with a non-trivial dilaton profile \cite{Alvares:2012kr} and look at the action that controls the vacuum configuration of the D7. Note that the dilaton will naturally be a function of the radial coordinate of the background space, $r=\sqrt{\rho^2 + L^2}$.  Now if we linearize (\ref{this}) in $L$ and make the coordinate transformation
\begin{equation} 
e^\phi \rho^3 {d \over d \rho} = \bar{\rho}^3 {d \over d \bar{\rho}}, \end{equation}
we obtain the action
\begin{equation} S \sim \int d \bar{\rho} \bar{\rho}^3 \left((\partial_{\bar{\rho}} L)^2 - {\Delta m^2 \over \bar{\rho}^2}  L\right), 
\end{equation} \newpage
with
\begin{equation} \Delta m^2 = - {\rho^5 \over \bar{\rho}} e^{\phi} {\partial e^\phi \over \partial \rho}. \end{equation}
The effect of a running coupling is precisely to introduce a radially dependent shift in the mass of the field $L$ as we have introduced in the Dynamic AdS/QCD model in (\ref{act}).

\end{document}